\documentclass[amsmath,amssymb,floatfix,prb,superscriptaddress,aps]{revtex4}

\usepackage{graphicx}
\usepackage{epsfig}
\usepackage{dcolumn}
\usepackage{bm}
\usepackage{times}

\begin{document}


\title{Duality between Poisson and Schr\"odinger equations in three dimensions}

\author{Gabriel Gonz\'alez}
\email{gabriel.gonzalez@uaslp.mx}
\affiliation{C\'atedra CONACYT-Coordinaci\'on para la Innovaci\'on y Aplicaci\'on de la Ciencia y la Tecnolog\'ia, \\ Universidad Aut\'onoma de San Luis Potos\'i, San Luis Potos\'i, 78000 MEXICO}

\date{\today}
\begin{abstract}
A duality between an electrostatic problem in a three dimensional world and a quantum mechanical problem in a one dimensional world which allows one to obtain the ground state solution of the Schr\"odinger equation by using electrostatic results is generalized to three dimensions. Here, it is demonstrated that the
same transformation technique is also applicable to the s-wave Schr\"odinger equation in three dimensions for central potentials. This approach leads to a general
relationship between the electrostatic potential and the s-wave function and the electric energy density to the quantum mechanical energy.
\end{abstract}
\maketitle

Recently in Ref.(\cite{Gonzalez, Gonzalez1, Gonzalez2}) a mapping between the electrostatic potential in three dimensions with various parallel charged planes and the one-dimensional Schr\"odinger equation was obtained through a simple transformation which allows for very simple analytical ground state wave functions and associated energies. This paper involve a planar symmetry which displays a direct relation with the one-dimensional Schr\"odinger equation. In the present article the transformation given in Ref. (\cite{Gonzalez}) is shown to work in the same straightforward fashion to the three dimensional case. Thus, with the same transformation, a student can directly obtain the ground state wave function and energy of more realistic potentials.\\
We first consider the transformation that maps the Poisson equation into a Schr\"odinger like equation. The Poisson equation in three dimensions is given by 
\begin{equation}
\nabla^2V=-\frac{\rho}{\epsilon_0},
\label{eq01}
\end{equation}
where $V$ is the electric potential, $\rho$ is the volume charge density and $\epsilon_0$ is the electric permitivity of free space.\cite{Griffiths, Gonzalez3,Gonzalez4,Gonzalez5, Gonzalez6} \\ Making the following transformation $V(\vec{r})=V_0\ln(\Psi(\vec{r})/A)$,\cite{Gonzalez7} where $V_0$ and $A$ are constants to ensure dimensional consistency, we obtain
\begin{equation}
V_0\frac{\nabla^2\Psi}{\Psi}-\frac{1}{V_0}\vec{E}\cdot\vec{E}=-\frac{\rho}{\epsilon_0},
\label{eq02}
\end{equation} 
where we have used the fact that $\vec{E}=-\nabla V=-V_0\frac{\nabla\Psi}{\Psi}$. Multiplying Eq. (\ref{eq02}) by $V_0\epsilon_0\Psi/2$ we end up with
\begin{equation}
-\frac{V_0^2\epsilon_0}{2}\nabla^2\Psi-\frac{1}{2}V_0\rho\Psi=-\frac{\epsilon_0}{2}\vec{E}\cdot\vec{E}\Psi
\label{eq02a}
\end{equation}
For spherically symmetric charge distributions the electrostatic potential depends on one variable only, i.e. $\Psi=\psi(r)$, and if the total electrostatic field is given by $\vec{E}=\vec{E}_1+\vec{E}_2$ where $\vec{E}_1$ is constant, then the total electrostatic energy density is given by ${\cal U}_{tot}=\epsilon_0|\vec{E}|^2/2={\cal U}_1+{\cal U}_2+\epsilon_0\vec{E}_1\cdot\vec{E}_2$. For this case Eq. (\ref{eq02a}) is given by 
\begin{equation}
-\frac{V_0^2\epsilon_0}{2r^2}\frac{d}{dr}\left(r^2\frac{d\psi}{dr}\right)+\left({\cal U}_2+\epsilon_0\vec{E}_1\cdot\vec{E}_2-\frac{1}{2}V_0\rho(r)\right)\psi=-{\cal U}_1\psi
\label{eq3}
\end{equation}
where $\psi(0)$ is a finite number and $\psi(r)\rightarrow 0$ as $r\rightarrow\infty$. 
Making the following substitutions:
\begin{equation}
V_0^2\epsilon_0a_0^3=\frac{\hbar^2}{m}, \quad \left({\cal U}_2+\epsilon_0\vec{E}_1\cdot\vec{E}_2-\frac{1}{2}V_0\rho(r)\right)a_0^3=U(r) \quad\mbox{and}\quad {\cal U}_1a_0^3=|E|
\label{eq3a}
\end{equation}
where $a_0$ is an arbitrary length we end up with the radial Schr\"odinger equation
\begin{equation}
-\frac{h^2}{2mr^2}\frac{d}{dr}\left(r^2\frac{d\psi}{dr}\right)+U(r)\psi=-|E|\psi
\label{eq3b}
\end{equation}
for bound states when $\ell=0$.\cite{ma, lau} 
As an example we are going to consider the charge distribution given by $\rho(r)=\sigma/r$, then the magnitude of the electrostatic field is obtained by using Gauss's law which is $|\vec{E}|=\sigma/2\epsilon_0$. It is interesting to note that the magnitude of the electrostatic field for this charge distribution is the same as the one produced by a infinite charge plane with uniform surface charge $\sigma$.\cite{Griffiths} Substituting the charge distribution and the magnitude of the electrostatic field into Eq.(\ref{eq3}) one has
\begin{equation}
-\frac{V_0^2\epsilon_0}{2r^2}\frac{d}{dr}\left(r^2\frac{d\psi}{dr}\right)-\frac{V_0\sigma}{2r}\psi=-\frac{\epsilon_0}{2}\left(\frac{\sigma}{2\epsilon_0}\right)^2\psi
\label{eq4}
\end{equation}
Using Eq.(\ref{eq3a}) and substituting in Eq.(\ref{eq4}) yields the following radial Schr\"odinger equation  
\begin{equation}
-\frac{h^2}{2mr^2}\frac{d}{dr}\left(r^2\frac{d\psi}{dr}\right)-\frac{\hbar\sigma}{2r}\psi=-\frac{m\sigma^2}{8}\psi
\label{eq5}
\end{equation}
It is convenient here to change the units of length and energy into atomic units. Hence if we put $e=m=\hbar=1$ in Eq. (\ref{eq5}) we have
\begin{equation}
-\frac{1}{2r^2}\frac{d}{dr}\left(r^2\frac{d\psi}{dr}\right)-\frac{\sigma}{2r}\psi=-\frac{\sigma^2}{8}\psi
\label{eq6}
\end{equation}
If we choose $\sigma=2Z$ where $Z$ denotes the atomic number then Eq.(\ref{eq6}) turns into 
\begin{equation}
-\frac{1}{2r^2}\frac{d}{dr}\left(r^2\frac{d\psi}{dr}\right)-\frac{Z}{r}\psi=-\frac{Z^2}{2}\psi
\label{eq7}
\end{equation}
Equation (\ref{eq7}) is the radial Schr\"odinger equation with a Coulomb interaction where the ground state wave function and the energy of Eq.(\ref{eq7}) are given by $\psi(r)=Ae^{-Zr}$ and $E=-\frac{Z^2}{2}$, respectively.\\
As a final example let us consider the case when $\rho(r)=\sigma/r+\rho_0$, then the total electrostatic field is obtained by using Gauss's law and the superposition principle which gives $|\vec{E}|=\sigma/2\epsilon_0+\rho_0r/3\epsilon_0$. Substituting the charge distribution and the magnitude of the electrostatic field into Eq.(\ref{eq3}) one has
\begin{equation}
-\frac{V_0^2\epsilon_0}{2r^2}\frac{d}{dr}\left(r^2\frac{d\psi}{dr}\right)+\left[\frac{\epsilon_0}{2}\left(\frac{\rho_0r}{3\epsilon_0}\right)^2+\frac{\sigma\rho_0r}{6\epsilon_0}-\left(\frac{V_0\sigma}{2r}+\frac{V_0\rho_0}{2}\right)\right]\psi=-\frac{\epsilon_0}{2}\left(\frac{\sigma}{2\epsilon_0}\right)^2\psi
\label{eq8}
\end{equation}
Using Eq.(\ref{eq3a}) and substituting in Eq.(\ref{eq8}) yields the following radial Schr\"odinger equation in atomic units 
\begin{equation}
-\frac{1}{2r^2}\frac{d}{dr}\left(r^2\frac{d\psi}{dr}\right)+\left[\frac{1}{2}\left(\frac{\rho_0r}{3}\right)^2+\frac{\sigma\rho_0r}{6}-\frac{\sigma}{2r}\right]\psi=-\left(\frac{\sigma^2}{8}-\frac{\rho_0}{2}\right)\psi
\label{eq9}
\end{equation}
If we choose $\sigma=2Z$ and $\rho_0=3W$ then Eq.(\ref{eq9}) turns into 
\begin{equation}
-\frac{1}{2r^2}\frac{d}{dr}\left(r^2\frac{d\psi}{dr}\right)+\left[\frac{1}{2}W^2r^2+ZWr-\frac{Z}{r}\right]\psi=-\left(\frac{Z^2}{2}-\frac{3}{2}W\right)\psi
\label{eq10}
\end{equation}
Equation (\ref{eq10}) is the radial Schr\"odinger equation with a Coulomb, linear, plus harmonic oscillator interaction where the ground state wave function and the energy of Eq.(\ref{eq10}) are given by $\psi(r)=Ae^{-Zr-Wr^2/2}$ and $E=-\frac{Z^2}{2}+\frac{3W}{2}$, respectively. It is interesting to note that there are bound state solutions only when $Z^2>3W$. In order to have physically realizable states we need to impose the following normalization condition 
\begin{equation}
\int_0^{\infty}r^2|\psi(r)|^2dr=1
\label{eq11}
\end{equation}
Applying Eq.(\ref{eq11}) to the ground state wave function $\psi(r)=Ae^{-Zr-Wr^2/2}$ we obtain the normalization constant for $Z>0$ and $W>0$ which is
\begin{equation}
A=\frac{2W^{5/2}}{\sqrt{-2Z\sqrt{W}+\sqrt{\pi}e^{Z^2/W}(W+2Z^2)Erfc[Z/\sqrt{W}]}}
\label{eq12}
\end{equation}
where $Erfc[Z/\sqrt{W}]$ is the complementary error function.\cite{leb}\\
By following this procedure the reader may be challenged to find exact ground state functions and energies for the radial Schr\"odinger equation for central potentials.

\end{document}